\documentstyle[epsfig]{elsart}
\newcommand{\newc}{\newcommand}
\newc{\lra}{\leftrightarrow}
\newc{\beq}{\begin{equation}}
\newc{\eeq}{\end{equation}}
\newc{\barr}{\begin{eqnarray}}
\newc{\earr}{\end{eqnarray}}

\newcommand{\beqa}{\begin{eqnarray}}
\newcommand{\eeqa}{\end{eqnarray}}
\newcommand{\bdm}{\begin{displaymath}}
\newcommand{\edm}{\end{displaymath}}

\renewcommand{\thefootnote}{\#\arabic{footnote}}
\renewcommand{\theequation}{\thesection.\arabic{equation}}
\renewcommand{\thefigure}{\thesection.\arabic{figure}}
\renewcommand{\thetable}{\thesection.\arabic{table}}

\begin{document}
%\begin{frontmatter}
\date{\today}
\title {A NETWORK OF NEUTRAL CURRENT SPHERICAL TPC'S  FOR DEDICATED
SUPERNOVA DETECTION}
%\title {EXPLOITING  TPC DETECTORS AND THE COHERENT  NEUTRAL CURRENT INTERACTION 
%FOR DETECTING  SUPERNOVA NEUTRINOS}
%
%
%\toctitle{ Neutrinos \protect\newline in a Spherical Box}
% allows explicit linebreak for the table of content
%
%
%\titlerunning{NEUTRINOS IN A SPHERICAL BOX}
% allows abbreviation of title, if the full title is too long
% to fit in the running head
%
\author{Y. Giomataris$^{1}$ and J.D. Vergados$^{2}$ }
%
%\authorrunning{Giomataris and Vergados}
% if there are more than two authors,
% please abbreviate author list for running head
%

\address{
1 CEA, Saclay, DAPNIA, Gif-sur-Yvette, Cedex,France}
\address{
2 University of Ioannina, Ioannina, GR 45110, Greece
\\E-mail:Vergados@cc.uoi.gr}
%\maketitle              % typesets the title of the contribution
\begin{frontmatter}
\begin{abstract}
The coherent contribution of all neutrons in neutrino nucleus scattering due to the neutral current 
offers a realistic prospect of detecting supernova neutrinos. As a matter of fact. for a 
typical supernova at $10$ kpc, about 1000 events are expected using 
 a spherical gaseous detector  of radius 4 m and
employing Xe gas at a pressure of 10 Atm. We propose a world wide network of several such simple,
stable and low cost supernova detectors with a running time of a few centuries.
\end{abstract}

\end{frontmatter}
\section{Introduction.}
The a typical supernova an energy of about $10^{53}$ ergs is released in the form of neutrinos
\cite{BEACFARVOG},\cite{SUPERNOVA}. These neutrinos
are emitted within an interval of about $10$ s after the explosion and they travel to Earth undistorted, except that,
on their way to Earth, they may oscillate into other flavors. The phenomenon of neutrino oscillations is by
 now established by the observation of
 atmospheric neutrino oscillations \cite{SUPERKAMIOKANDE} interpreted as
 $\nu_{\mu} \rightarrow \nu_{\tau}$ oscillations, as well as
 $\nu_e$ disappearance in solar neutrinos \cite{SOLAROSC}. These
 results have been recently confirmed by the KamLAND experiment \cite{KAMLAND},
 which exhibits evidence for reactor antineutrino disappearance. Thus for traditional detectors
 relying on the charged current
 interactions the precise event rate may depend critically on the specific properties of the supernova, in particular
 its distance from the Earth. This, of course, may turn into an advantage for the study of the neutrino properties.
 An additional problem is the fact that the charged current cross sections depend on the details of the 
 structure of the nuclei involved. 
 
 In recent years, however, it has become feasible to detect neutrinos by measuring
 the recoiling nucleus employing gaseous detectors. Thus
 one is able to explore the advantages offered by the neutral current interaction. This way
 there are no problems associated with uncertainties in the flux of any neutrino flavor due to oscillations.
Furthermore this interaction, through its vector component, can lead
 to coherence, i.e. an additive contribution of all nucleons in the nucleus. Since the vector  contribution of the
 protons is tiny, the coherence is mainly due to the neutrons of the nucleus.
 
 In this paper we will derive the amplitude for the differential neutrino nucleus coherent cross section. Then
 we will utilize the available information regarding the energy spectrum of supernova neutrinos and evaluate
 the expected number of events for all the noble gas targets. Then we will show that these results can be exploited
by a network of small and relatively cheap spherical TPC detectors placed in various parts of the world 
(for a description of the apparatus see our
earlier work \cite{NOSTOS1}). The operation of such devices as a network will minimize the background 
problems. There is no need to go underground,
but one may have to go sufficiently deep underwater to balance the high pressure of the gas target. Other types
of detectors have also been proposed \cite{MICROPATTERN},\cite{TWOPHASE}.\\
Large gaseous volumes are easily obtained by employing long drift technology (i.e TPC) that can 
provide massive targets by increasing the gas pressure. Combined with an adequate amplifying 
structure and low energy thresholds, a three-dimensional reconstruction of the recoiling particle, electron
 or nucleus, can be obtained. The use of new micropattern detectors and especially the novel Micromegas 
\cite{GIOMA96} provide excellent spatial and time accuracy that is a precious tool for pattern recognition and
 background rejection \cite{CG01},\cite{GORO05}.
The virtue of using such large gaseous volumes and the new high precision microstrip gaseous detectors 
has been recently discussed in a dedicated workshop \cite{WORKSHOP04} and their relevance for low energy neutrino 
physics and dark matter detection has been widely recognized. Such  low-background low-energy threshold 
systems are actually successfully used in the CAST \cite{AALSETH} solar axion experiment 
and are under development 
for several low energy neutrino or dark matter projects \cite{NOSTOS1},\cite{DARKMATTER}.
\section{ Elastic Neutrino nucleon Scattering}
\begin{figure}[!ht]
 \begin{center}
%\rotatebox{90}{\hspace{0.0cm} {$\rightarrow \frac{T_{recoil}}{m_{recoil}}$}}
%\includegraphics[scale=0.3]{scale_ener1.eps}
%\hspace{1.0cm}$\rightarrow \frac{E_{\nu}}{m_{recoil}}$
 \rotatebox{90}{\hspace{0.0cm} {$\rightarrow \frac{T_{recoil}}{m_{recooil}}$}}
\includegraphics[scale=0.6]{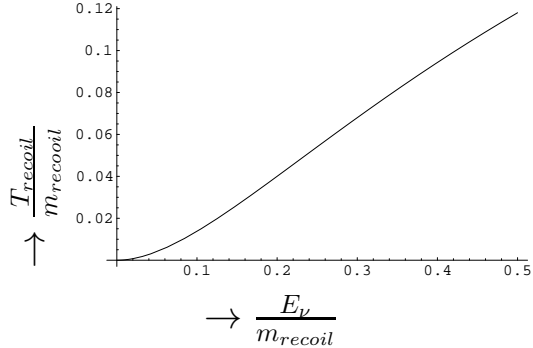}
\hspace{6.0cm}$\rightarrow \frac{E_{\nu}}{m_{recoil}}$
 \caption{The maximum recoil energy  as a function of the neutrino energy (both in units of the
recoiling mass)}
 \label{fig:yofx}
 \end{center}
  \end{figure}

\begin{figure}[!ht]
 \begin{center}
 \rotatebox{90}{\hspace{1.0cm} {$\rightarrow \frac{E_{\nu}}{m_{recoil}}$}}
\includegraphics[scale=0.8]{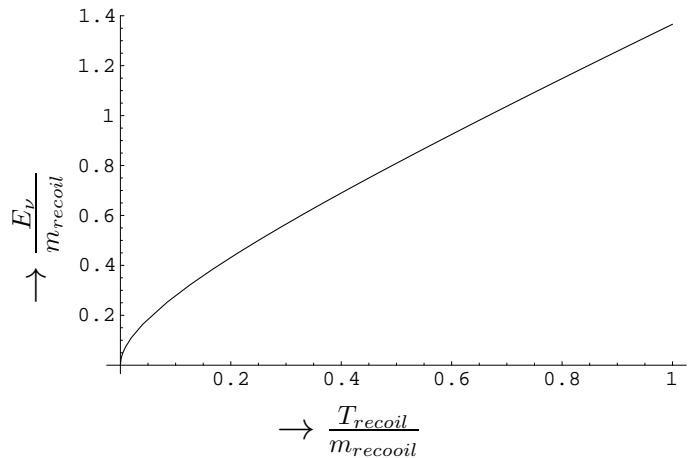}
\hspace{6.0cm}$\rightarrow \frac{T_{recoil}}{m_{recooil}}$
% \rotatebox{90}{\hspace{1.0cm} {$\rightarrow \frac{T_{recoil}}{m_{recooil}}$}}
%\includegraphics[scale=0.4]{scale_ener2.eps}
%\hspace{1.0cm}$\Rightarrow \frac{E_{\nu}}{m_{recoil}}$
 \caption{The minimum neutrino energy as a function of the recoil energy (both in units of the
recoiling mass)}
 \label{fig:xofy}
 \end{center}
  \end{figure}

% For low energy neutrinos the historic process neutrino-electron scattering \cite{HOOFT} \cite{REINES}
% is very useful.
%The differential cross section \cite{VogEng} takes the form
%('t Hooft and Vogel $\&$ Engel)
%\begin{equation}
%\frac{d\sigma}{dT}=\left(\frac{d\sigma}{dT}\right)_{weak}+
%\left(\frac{d\sigma}{dT}\right)_{EM} \label{elas1a}
%\end{equation}
The cross section for elastic neutrino nucleon scattering has extensively been studied.
% \cite{ELNUNUC}.
It has been shown that at low energies it can be simplified and  be cast in the form:
\cite{BEACFARVOG},\cite{VogEng}:
 \begin{eqnarray}
 \left(\frac{d\sigma}{dT_N}\right)_{weak}&=&\frac{G^2_F m_N}{2 \pi}
 [(g_V+g_A)^2\\
\nonumber
&+& (g_V-g_A)^2 [1-\frac{T_N}{E_{\nu}}]^2
+ (g_A^2-g_V^2)\frac{m_NT_N}{E^2_{\nu}}]
 \label{elasw}
  \end{eqnarray}
  where $m_N$ is the nucleon mass and $g_V$, $g_A$ are the weak coupling constants. Neglecting their
  dependence on the momentum transfer to the nucleon they take the form:
  \beq
 g_V=-2\sin^2\theta_W+1/2\approx 0.04~,~g_A=\frac{1.27}{2} ~~,~~(\nu,p)
\label{gcoup1}
\eeq
\beq
g_V=-1/2~,~g_A=-\frac{1.27}{2}~~,~~(\nu,n)
 \label{gcoup2}
 \eeq
 In the above expressions for the axial current the renormalzation
in going from the quark to the nucleon level was taken into account. For antineutrinos $g_A\rightarrow-g_A$.
To set the scale we write:
\beq \frac{G^2_F m_N}{2 \pi}=5.14\times 10^{-41}~\frac{cm^2}{MeV}
\label{weekval} 
\eeq
% In the above expressions for the $\nu_{\mu},\nu_{\tau}$ only the
% neutral current has been included, while for $\nu_e$ both the
% neutral and the charged current contribute.
The nucleon energy depends on the  neutrino energy and the
scattering angle and is given by:

$$T_N= \frac{2~m_N (E_{\nu}\cos{\theta})^2}{(m_N+E_\nu)^2-(E_{\nu}
\cos{\theta})^2}$$
%$$T=\frac{X^2}{2 m_e}~~,~~X=2E_{\nu} \frac{m_e(m_e+E_{\nu})\cos{\theta}}
%{(m_e+E_\nu)^2-(E_{\nu} \cos{\theta})^2}$$
The last equation can be simplified as follows:
 $$T_N \approx \frac{ 2(E_\nu \cos{\theta})^2}{m_N}$$
The above formula can be generalized to any target of mass m. It can be written in dimensionless form
as follows:
\beq
y=\frac{2\cos^2{\theta}}{(1+1/x)^2-\cos^2{\theta}}~~,~~y=\frac{T}{m},x=\frac{E_{\nu}}{m}
\label{recoilen}
\eeq 
 The maximum  energy occurs when $\theta=0$ and
 depends on the neutrino energy (see Fig. \ref{fig:yofx}).
%I what follows, whenever appropriate, we are going to average our results with the neutrino spectrum 
%shown in Fig. \ref{spectrum}.
  One can invert the above equation and get the minimum neutrino energy associated with a given recoil energy.
  This is useful in obtaining the differential cross section (with respect to the recoil energy) after folding
  with the neutrino spectrum. One finds:
  \beq
  x=\left[-1+\sqrt{1+\frac{2}{y}} \right]^{-1}
  \label{xofy}
  \eeq
  This is shown in Fig. \ref{fig:xofy}

\section {Coherent neutrino nucleus scattering}
\begin{figure}[!ht]
 \begin{center}
 \rotatebox{90}{\hspace{1.0cm} {$F_{coh}$}}
\includegraphics[scale=0.6]{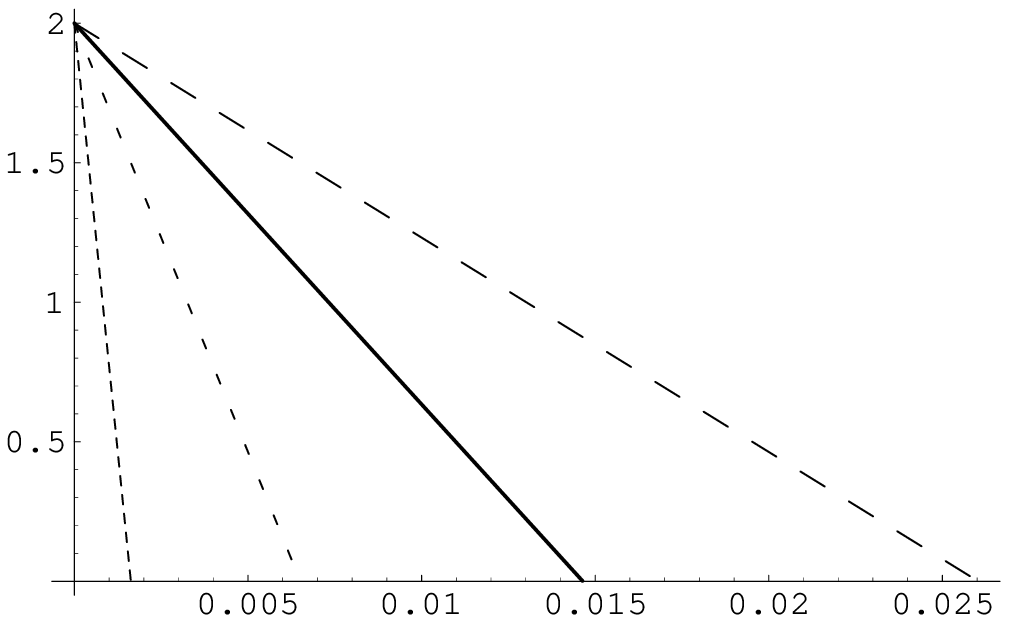}
\includegraphics[scale=0.6]{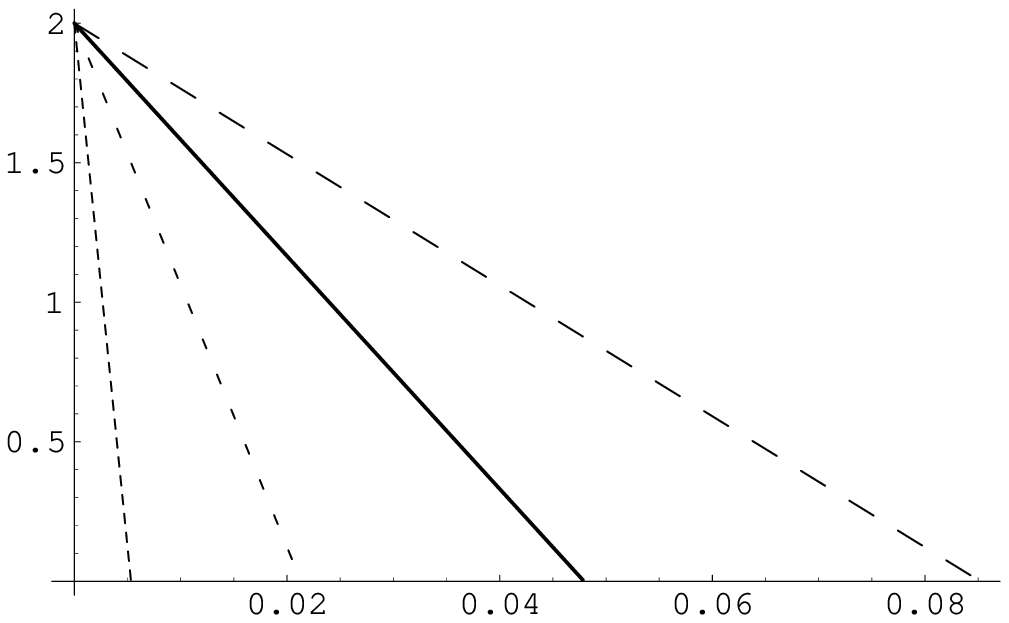}
\hspace{8.0cm}$T_A \rightarrow$ MeV
 \caption{The function $F_{coh}(A,T_A,E_{\nu})$ as a function of the recoil energy $T_A$ for,
from left to right, $E_{\nu}=10,~20,~30,~40$
MeV. The results shown are for Xe on the left and Ar on the right}
 \label{fig:fcoh}
 \end{center}
  \end{figure}

From the above expressions we see that the vector current contribution, which may lead to coherence, is negligible
in the case of the protons. Thus the coherent contribution \cite{PASCHOS} may come from the neutrons and is expected to be
proportional to the square of the neutron number.
The neutrino-nucleus scattering can be obtained from the amplitude of the neutrino nucleon scattering under
the following assumptions:
\begin{itemize}
\item Employ the appropriate kinematics, i.e. those involving the elastically scattered nucleus.
\item Ignore  effects of the nuclear form factor. Such effects, which are not expected to be very large,
 are currently under study and they will
 appear elsewhere.
\item The effective neutrino-nucleon amplitude is obtained as above with the substitution 
$${\bf q}\Rightarrow \frac{{\bf p}}{A}~~,~~E_N \Rightarrow \sqrt{m_N^2+\frac{{\bf p}^2}{A^2}}=\frac{E_A}{A}$$
with ${\bf q}$ the nucleon momentum and ${\bf p}$ the nuclear momentum.  
\end{itemize}
Under the above assumptions the neutrino-nucleus cross section takes the form:
 \begin{eqnarray}
 \left(\frac{d\sigma}{dT_A}\right)_{weak}&=&\frac{G^2_F Am_N}{2 \pi}
 [(M_V+M_A)^2 \left (1+\frac{A-1}{A}\frac{T_A}{E_{\nu}} \right )
 \nonumber\\
+ (M_V-M_A)^2 
(1&-&\frac{T_A}{E_{\nu}})^2
\left (1-\frac{A-1}{A}\frac{T_A}{m_N}\frac{1}{E_{\nu}/T_A-1} \right )
\nonumber\\
&+& (M_A^2-M_V^2)\frac{Am_NT_A}{E^2_{\nu}} ]
 \label{elaswA}
  \end{eqnarray}
  Where $M_V$ and $M_A$ are the nuclear matrix elements associated with the vector and the axial current
  respectively and $T_A$ is the energy of the recoiling nucleus.
 The axial current contribution vanishes for $0^+ \Rightarrow 0^+$ transitions. Anyway it is negligible
  in front of the coherent scattering due to neutrons. Thus the previous formula is reduced to:
   \begin{eqnarray}
 \left(\frac{d\sigma}{dT_A}\right)_{weak}&=&\frac{G^2_F Am_N}{2 \pi}~(N^2/4)F_{coh}(A,T_A,E_{\nu}),
\nonumber\\
& &F_{coh}(A,T_A,E_{\nu})=
  \left (1+\frac{A-1}{A}\frac{T_A}{E_{\nu}} \right )
+(1-\frac{T_A}{E_{\nu}})^2
\nonumber\\
& &\left (1-\frac{A-1}{A}\frac{T_A}{m_N}\frac{1}{E_{\nu}/T_A-1} \right )
-\frac{Am_NT_A}{E^2_{\nu}} 
 \label{elaswAV}
  \end{eqnarray}
  The function $F_{coh}(A,T_A,E_{\nu})$ is shown in Fig \ref{fig:fcoh} as a function of the recoil
  energy in the case of Ar and Xe for $10,20,30$ and $40$ MeV respectively.
  
  We see two reasons for enhancement of the cross section: 
\begin{itemize}
\item The overall A factor due to the kinematics, which is  counteracted by the smaller
nuclear recoil energy when compared to the nucleon recoil energy for the same neutrino energy. 
This factor will be absorbed into the energy integrals, see the function $F_{fold}(A,T,(T_A)_{th})$ below.
\item The $N^2$ enhancement due to coherence.
\end{itemize}
\section{Supernova Neutrinos}
\begin{figure}[!ht]
 \begin{center}
 \rotatebox{90}{\hspace{1.0cm} {$\frac{dN}{d E_{\nu}}\rightarrow \frac{10^{58}}{MeV}$}}
\includegraphics[scale=0.8]{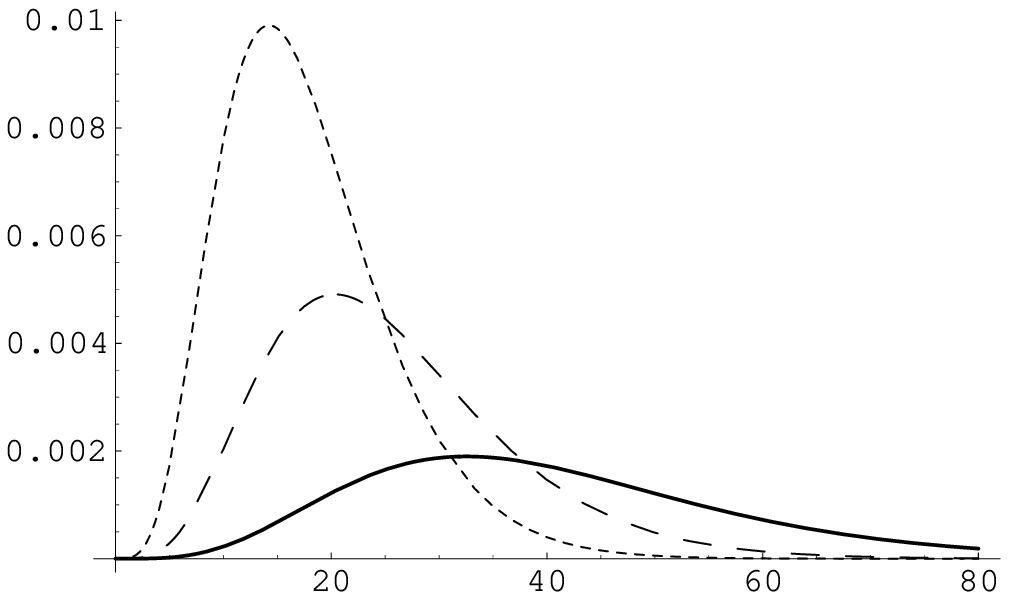}
\hspace{8.0cm}$E_{\nu} \rightarrow$ MeV
 \caption{The supernova neutrino spectrum. The short dash, long dash and continuous curve correspond
to $\nu_e,\tilde{\nu}_e$ and all other flavors respectively}
 \label{supernovasp}
 \end{center}
  \end{figure}

The number of neutrino events for a given detector depends on the neutrino spectrum and the distance of the
source. We will consider a typical case of a source which is about $10$ kpc, l.e. $D=3.1 \times 10^{22}$ cm with 
an energy output of $3 \times 10^{53}$ ergs with a duration of about $10$ s. We will further assume that the energy
is shared equally by each neutrino flavor. Furthermore each neutrino flavor is characterized by  a 
 Fermi-Dirac like distribution times its characteristic cross section, i.e $U_{\nu}=0.5 \times 10^{53}$ ergs
per neutrino flavor, i.e.
\beq
\frac{dN}{dE_{\nu}}=\sigma(E_{\nu})\frac{E^2_{\nu}}{1+exp(E_{\nu}/T)}=\frac{\Lambda}{JT}\frac{x^4}{1+e^x}
\label{nudistr}
\eeq
with
$J=\frac{31\pi^6}{252}$, $\Lambda$  a constant and 
$T$ the temperature of the emitted neutrino flavor. 
Each flavor is characterized by its
own temperature as follows:
$$T=8 \mbox { MeV for } \nu_{\mu},\nu_{\tau},\tilde{\nu}_{\mu}, \tilde{\nu}_{\tau}
\mbox{ and } T=5 ~(3.5)\mbox{ MeV for } \tilde{\nu}_e ~(\nu_e)$$
The constant $\Lambda$ is determined by the requirement that the distribution yields the total energy of each
neutrino species.
$$U_{\nu}=\frac{\Lambda T}{J}\int_0^{\infty } dx \frac{x^5}{1+e^x}\Rightarrow \Lambda=\frac{U_{\nu}}{T}$$
Thus one finds:
$$\Lambda=0.89\times 10^{58}~(\nu_e),~~0.63\times 10^{58}~(\tilde{\nu}_e)~,0.39\times 10^{58}
\mbox{ (all other flavors)}$$
The emitted neutrino spectrum is shown in Fig. \ref{supernovasp}.
  
\begin{figure}[!ht]
 \begin{center}
% \rotatebox{90}{\hspace{1.0cm} {$F_{coh}$}}
\includegraphics[scale=0.6]{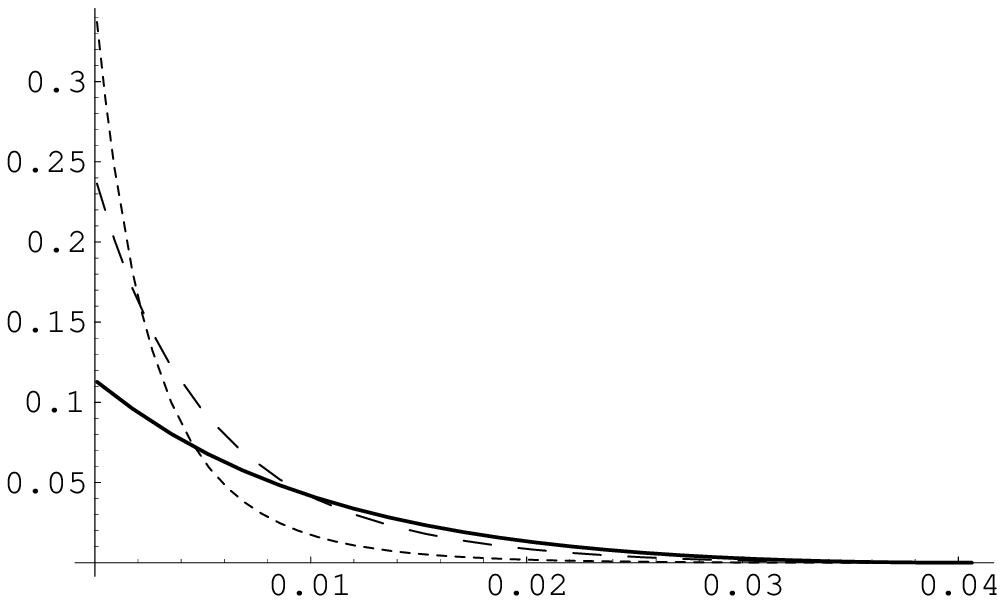}
\includegraphics[scale=0.6]{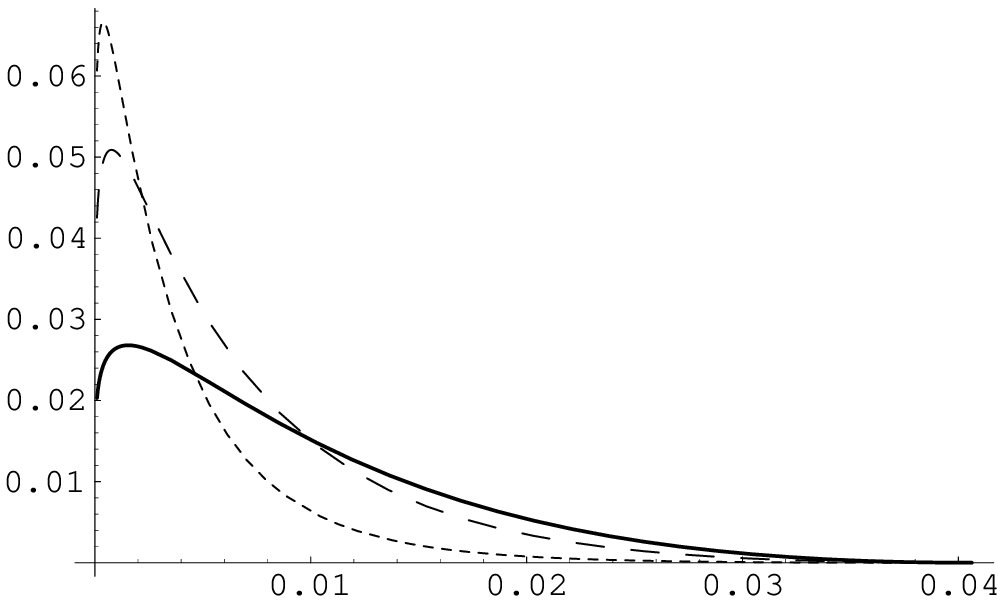}
\hspace{8.0cm}$T_A \rightarrow$ MeV
 \caption{The differential event rate as a function of the recoil energy $T_A$, in arbitrary units, for
Xe. On  the left we show the results without quenching, while on the right the quenching factor is
included. We notice that the effect of quenching is more prevalent at low energies. The notation 
for each neutrino species is 
the same as in Fig. \ref{supernovasp}}
 \label{fig:difr131}
 \end{center}
  \end{figure}

  The differential event rate (with respect to the recoil energy) is proportional to the quantity:
\beq
\frac{dR}{dT_A}=\frac{\lambda (T)}{J}\int_0^{\infty } dx
F_{coh}(A,T_A,xT) \frac{x^4}{1+e^x}
\label{dRdT}
\eeq
with $\lambda(T)=(0.89,0.63,0.39)$  for $\nu_e,\tilde{\nu}_e$ and all other flavors respectively. 
This is shown  in Figs. \ref{fig:difr131} and \ref{fig:difr131}.
   \begin{figure}[!ht]
 \begin{center}
% \rotatebox{90}{\hspace{1.0cm} {$F_{coh}$}}
\includegraphics[scale=0.6]{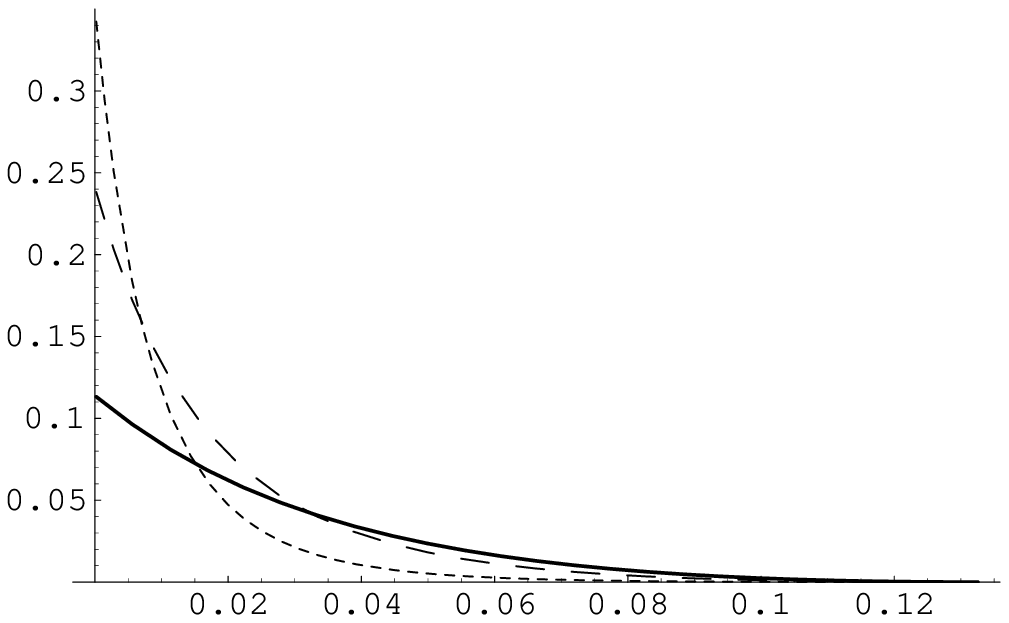}
\includegraphics[scale=0.6]{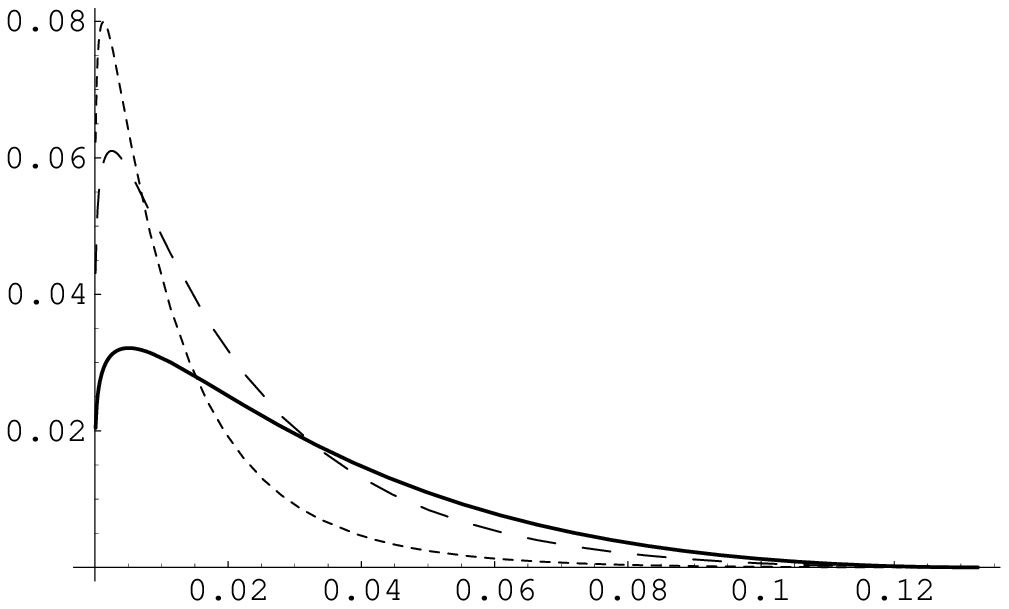}
\hspace{8.0cm}$T_A \rightarrow$ MeV
 \caption{The same as in Fig. \ref{fig:difr131} for the Ar target.}
 \label{fig:difr40}
 \end{center}
  \end{figure}
The total number of expected events for each neutrino species can be cast in the form:
\begin{eqnarray}
\mbox{No of events}&=&\tilde{C}_{\nu} (T) h(A,T,(T_A)_{th}),
\nonumber\\
\\h(A,T,(T_A)_{th})
&=&\frac{F_{fold}(A,T,(T_A)_{th})}{F_{fold}(40,T,(T_A)_{th})}
\label{events}
\end{eqnarray}
with
\begin{eqnarray}
F_{fold}(A,T,(T_A)_{th})&=&\frac{A}{J}\int_{(T_A)_{th}}^{(T_A)_{max}}\frac{dT_A}{1MeV}\times 
\nonumber\\ 
& &\int_0^{\infty } dx F_{coh}(A,T_A,xT) \frac{x^4}{1+e^x}
\label{events1}
\end{eqnarray}
and
\beq
\tilde{C}_{\nu}(T)=\frac{G^2_F m_N1MeV}{2 \pi} \frac{N^2}{4}\Lambda (T)\frac{1}{4 \pi D^2}\frac{PV}{kT_0}
\label{C1}
\eeq
Where $k$ is Boltzmann's constant, $P$ the pressure, $V$ the volume, and $T_0$ the temperature of the gas.

Summing over all the neutrino species we can write:
\beq
\mbox{No of events}=C_{\nu} r(A)\frac{K(A,(T_A)_{th})}{K(40,(T_A)_{th})}Qu(A)
\label{sumevents}
\eeq
with
\beq
C_{\nu}=153  \left ( \frac{N}{22} \right )^2 \frac{U_{\nu}}{0.5\times 10^{53}ergs}
\left ( \frac{10kpc}{D}\right )^2 \frac{P}{10Atm}
\left[ \frac{R}{4m}\right]^3 \frac{300}{T_0}
\label{C2}
\eeq
%with 
%\beq
%C=3.7 \times 10^{3}~(\nu_e),2.6\times 10^{3}~(\tilde{\nu}_e)~,1.6\times 10^{3}\mbox{ (other flavors)}
%\label{C3}
%\eeq
%The function $F_{fold}(A)$ scales as $A^{-1}$ so it is more convenient to write:
%\beq
%\mbox{No of events}=C_{\nu}r(A)~,~r(A)=\frac{h(40)}{h(A)}
%\label{events2}
%\eeq
%with
%\beq
%h(A)=\left[ F_{fold}(A) \right ]^{-1}
%\label{events3}
%\eeq
%The function $h(A,(T_A)_{th})$ can be written as follows:
%\beq
%h(A,(T_A)_{th})=r(A)K(A,(T_A)_{th})
%\label{factor}
%\eeq
In the above expression $r(A)$ is a kinematical parameter depending on the nuclear mass number,
which is essentially unity.

\begin{figure}[!ht]
 \begin{center}
 \rotatebox{90}{\hspace{-2.0cm} {$r(A)\rightarrow $}}
 \hspace{8.0cm}$A \rightarrow$ 
\includegraphics[scale=0.8]{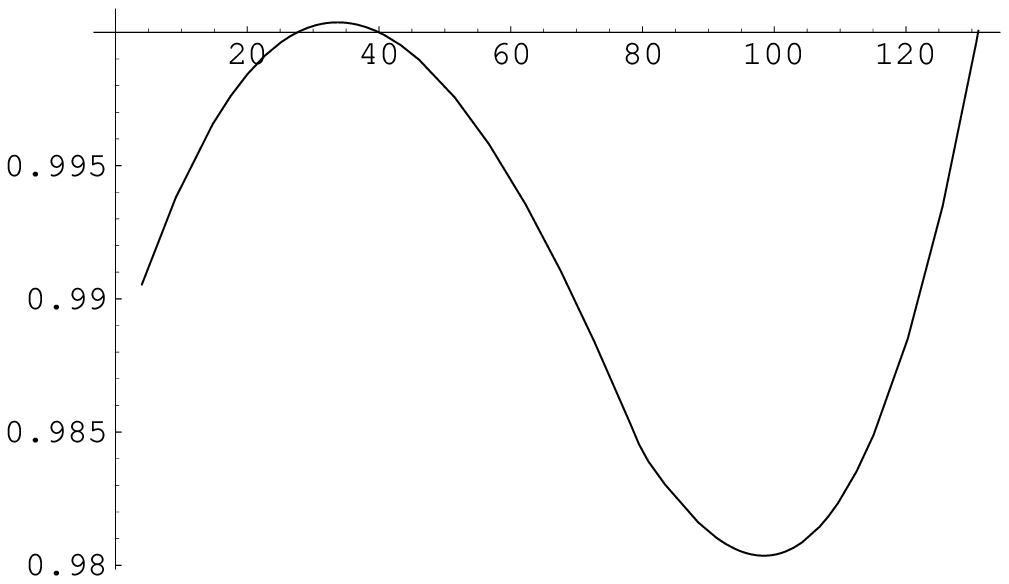}
%\hspace{8.0cm}$A \rightarrow$ 
 \caption{The function $r(A)$ versus the nuclear mass number. To a good approximation $r(A)\simeq 1.0$
(for definitions see text)}
 \label{fig:ratio}
 \end{center}
  \end{figure}

\begin{figure}[!ht]
 \begin{center}
 \rotatebox{90}{\hspace{-0.0cm} {${\tiny K(A,(T_A)_{th})}\rightarrow $}}
 %\hspace{0.0cm}${\tiny K(T_A)_{th}} \rightarrow$MeV 
\includegraphics[scale=0.8]{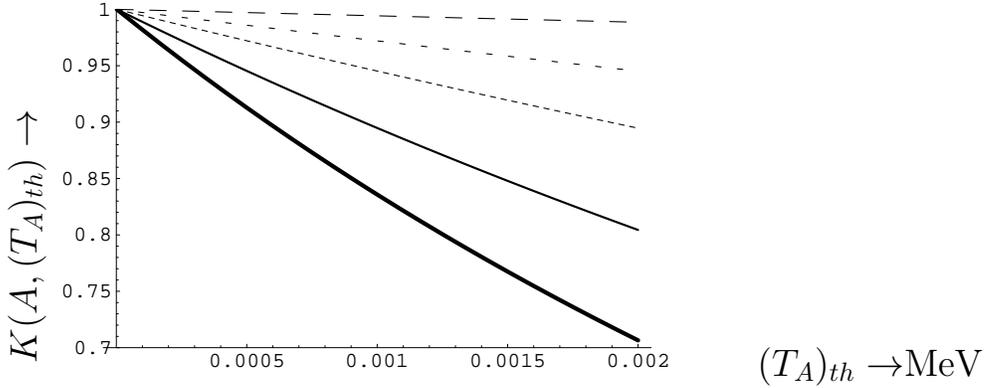}
 \hspace{1.0cm}${\tiny (T_A)_{th}} \rightarrow$MeV 
%  \vspace{0.5cm}
%  \rotatebox{90}{\hspace{-0.0cm} {${\tiny K(A,(T_A)_{th})}\rightarrow $}}
 %\hspace{0.0cm}${\tiny K(T_A)_{th}} \rightarrow$MeV 
%\includegraphics[scale=0.8]{th20.eps}
% \hspace{1.0cm}${\tiny K(T_A)_{th}} \rightarrow$MeV
% \vspace{0.5cm} 
% \rotatebox{90}{\hspace{-0.0cm} {${\tiny K(A,(T_A)_{th})}\rightarrow $}}
 %\hspace{0.0cm}${\tiny K(T_A)_{th}} \rightarrow$MeV 
%\includegraphics[scale=0.8]{th40.eps}
% \hspace{1.0cm}${\tiny K(T_A)_{th}} \rightarrow$MeV 
% \vspace{1.0cm}
%\rotatebox{90}{\hspace{-3.0cm} {$K(A,(T_A)_{th})\rightarrow $}}
 %\hspace{0.0cm}$K(T_A)_{th} \rightarrow$MeV 
%\includegraphics[scale=0.4]{th20.eps}
%\hspace{8.0cm}$A \rightarrow$ 
 \caption{The function $K(A,(T_A)_{th})$ versus $(T_A)_{th}$ for various the nuclear mass numbers
without the quenching factor.
From  top to bottom He, Ne, Ar, Kr and Xe.
(for definitions see text)}
 \label{fig:K}
 \end{center}
  \end{figure}

\begin{figure}[!ht]
 \begin{center}
 \rotatebox{90}{\hspace{-0.0cm} {${\tiny K(A,(T_A)_{th})}\rightarrow $}}
 %\hspace{0.0cm}${\tiny K(T_A)_{th}} \rightarrow$MeV 
\includegraphics[scale=0.6]{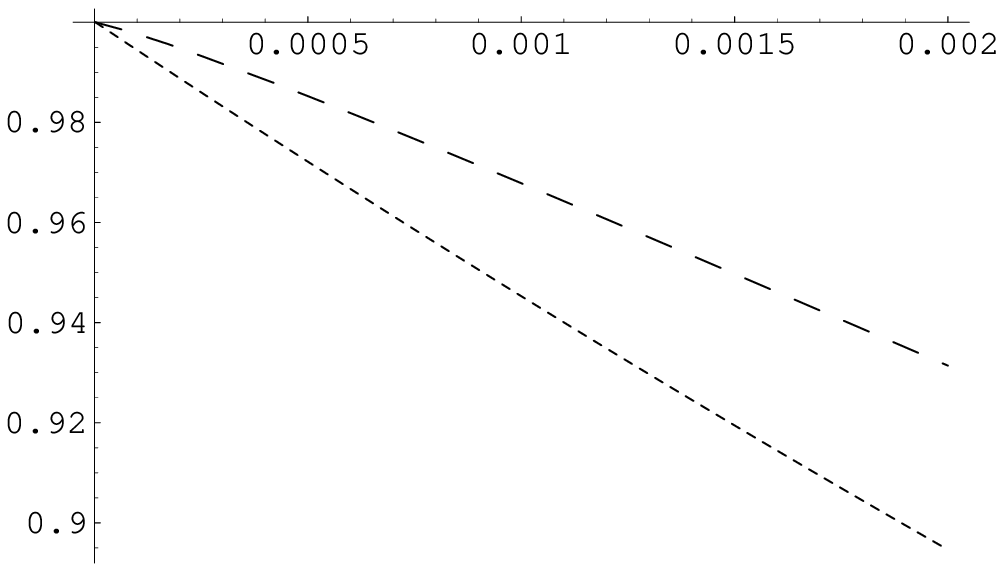}
\includegraphics[scale=0.6]{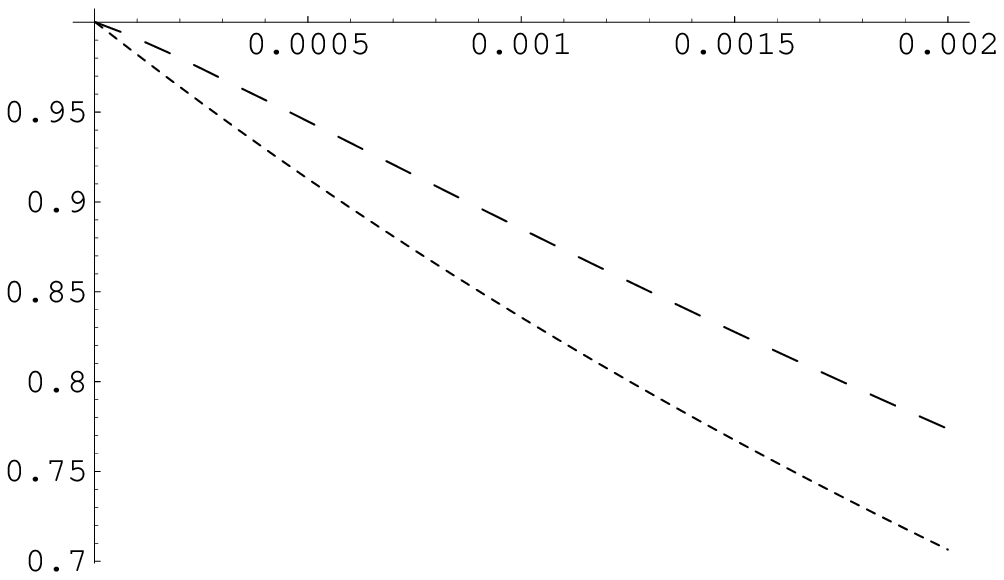}
 \hspace{1.0cm}${\tiny (T_A)_{th}} \rightarrow$MeV 
 \caption{The function $K(A,(T_A)_{th})$ versus $(T_A)_{th}$ for the target Ar on the left and
Xe on the right. The short and long dash correspond to no quenching and quenching factor respectively.
One sees that the effect of quenching is less pronounced at higher thresholds.
The differences appear small, since we present here only  the ratio of the rates to that at zero
threshold. The effect of quenching at some specific threshold energy is not shown here.
For a threshold energy of $100$ eV the rates are quenched by factors of $3$ and $3.5$ for Ar and Xe
 respectively (see Eq. (\ref{quench2}).}
 \label{fig:Kqu}
 \end{center}
  \end{figure}

% $(T_A)_{th}=0$, while
 $K(A,(T_A)_{th})$
 is  the rate at a given threshold energy diveded by that at zero threshold. It depends
 on the threshold
energy, the assumed quenching factor and the nuclear mass number. It is unity at $(T_A)_{th})=0$.
The function $r(A)$ is plotted in \ref{fig:ratio}. It is seen that  it can be well approximated by unity.\\
 From the above equation we find
that, ignoring quenching, the following expected number of events:
\beq
1.25,~31.6,~153,~614,~1880\mbox{ for He, Ne, Ar, Kr and Xe}
\label{allrates}
\eeq
respectively. For other possible targets the rates can be found by the above formulas or interpolation.\\
  The function $K(A,(T_A)_{th})$ is plotted in Fig.
 %\ref{fig:K1} and \ref{fig:K2}
\ref{fig:K} for threshold energies up to $2$keV. 
  We see that the threshold effects are stronger in heavier systems since, on the average, the transfered
  energy is smaller. Thus for a threshold energy of $2$ keV in the case of Xe the number of events is reduced
  by $30\%$ compared to those at zero threshold.\\
 The quantity $Qu(A)$ is a factor less than one multiplying the total rate, assuming a  threshold energy
  $(T_A)_{th}=100$eV, due to the quenching. The idea of quenching is introduced, since, for low emery recoils,
 only a fraction of the total deposited energy goes into
 ionization. The ratio of the amount of ionization induced in the gas due to nuclear recoil to the amount of ionization
 induced 
by an electron of the same kinetic energy is referred to as a quenching factor $Q_{fac}$. This factor depends mainly on the 
detector material, the recoiling energy as well as the process considered \cite{SIMON03}.
 In our estimate of $Qu(T_A)$ we assumed a quenching factor of the following empirical form motivated by the Lidhard
theory \cite{SIMON03}-\cite{LIDHART}:
\beq
Q_{fac}(T_A)=r_1\left[ \frac{T_A}{1keV}\right]^{r_2},~~r_1\simeq 0.256~~,~~r_2\simeq 0.153 
\label{quench1}
\eeq 
Then the parameter $Qu(A)$ takes the values:
\beq
0.49,~0.38,~0.35,~0.31,~0.29\mbox{ for He, Ne, Ar, Kr and Xe}
\label{quench2}
\eeq
respectively. The effect of quenching is larger in the case of  heavy targets, since, for a given neutrino energy, the energy of
the recoiling nucleus is smaller. Thus the number of expected events for Xe assuming a threshold energy of
$100$ eV is reduced to about 560.\\
  The effect of quenching is exhibited in Fig \ref{fig:Kqu}  for the two interesting targets 
Ar and Xe.
 \\We should mention that it is of paramount importance to experimentally measure the quenching factor. The
 above estimates were based on the assumption of a pure gas. In our detection scheme the Xe gas carrier
 (A) is mixed with a small fraction of low ionization potential gas (B). Thus a part of the excitations
 produced on the Xe atoms could be transferred to ionization through the well known Penning effect
 as follows:
 \beq
 A^*+B\longrightarrow A+B^{*+}+e^{-}
 \label{penning}
 \eeq 
 Such an effect will lead to an increase  in the quenching factor and needs be measured.
\section{The NOSTOS detector network}
A description of the NOSTOS project and details of the spherical TPC detector are given in \cite{NOSTOS1}.
 We have built a spherical prototype 1.3 m in diameter 
which is described in \cite{NOSTOS2}. The outer vessel is made of pure Cu (6 mm thick)
allowing to sustain pressures up to 5 bar. The inner detector is just a small sphere, 10 mm 
in diameter, made of stainless
steel as a proportional counter located at the center of
curvature of the TPC. We intend to use as amplifying structure
a spherical TPC \cite{GRRC} and  developments are currently under way  
to build a spherical TPC detector using new
technologies. First tests were performed by filling
the volume with argon mixtures and are quite promising. High gains are easily obtained and
 the signal to noise is large enough for sub-keV threshold. The whole system looks stable
 and robust.
The advantages of using the spherical detector concept are the following;
\begin{enumerate}
\item The natural radial focusing of the TPC allows to collect and amplify the deposited
 charges by a simple and robust detector using a single electronic channel to read out
 a large gaseous volume. The small size associated to small detector capacitance permits one
to achieve very low electronic noise. In the present prototype the noise is as low as a 
few hundred electrons and has  easily been obtained; with optimized low noise amplifiers we 
hope to lower it to the level of a few tenths. This is a key point for the obtaining a very
 low energy threshold, i.e. down to 100 eV, by operating the detector at moderate gain of about 100.
 Such low gains are easily obtained at atmospheric pressure and open the way to operate the
 TPC at high pressures. We target pressures as high as 10 bar for Xenon gas. Even higher pressures
 by a factor 3-6  are aimed in the case of Argon gas in order to achieve, to first order, the same number
 of events.
\item The radial electric field, inversely proportional to the square of the radius, is a crucial 
point for measuring the depth of the interaction by a simple analysis of the time structure of 
the detector signal. A position resolution of about 10 cm has been already obtained, a fact that is
 of paramount importance for improving the time resolution of the detector and rejecting
 background events by applying fiducial cuts.
\item Building a high pressure metallic sphere, for instance made out of stainless steel or 
copper, seems to assure an excellent quality of the gas mixture and turns out that a single 
gas filling with pure gas is sufficient to maintain the stability of the signal for several 
months. We are pushing the technology to improve the properties of the various elements in
 order to achieve stability over many years.
\item Big high pressure-secure tanks are under development by many international companies
 for hydrogen or oil storage, and therefore the main element of the TPC could be shipped 
at moderate cost.
\end{enumerate}
Our idea is then to  build several such low cost and robust detectors and install them in 
several places over the world.  First estimations show that the required background level
is modest and therefore there is not need for deep underground laboratory. A mere  100 meter 
water equivalent
coverage seems to be sufficient to reduce the cosmic muon flux at the required level 
(in the case of many such detectors in coincidence, a modest shield is sufficient). 
The maintenance of such system could be easily assured by Universities or even by 
secondary schools. Thanks to the simplicity of the system it could be operated by young 
students with a specific running program and simple maintenance every a few years. 
Notice that such detector scheme, measuring low energy nuclear recoils from neutrino 
nucleus elastic scattering, do not determine the incident neutrino vector and therefore 
it is not possible this way to localize the Supernova. A cluster of such detectors in 
coincidence, however, could localize the star by a triangulation technique. \\
 A network of such detectors in coincidence with a sub-keV threshold could also be used o observe
 unexpected low energy events. This low energy range has never been explored using massive
 detectors. A challenge of great importance will be the synchronization of such a detector cluster
with the astronomical $\gamma$-ray burst telescopes to establish whether low energy recoils are
emitted in coincidence with the mysterious $\gamma$ bursts.

\section{Conclusions}
 In the present study it has been shown that it is quite simple to detect typical supernova
neutrinos in our galaxy. The idea is to employ a small size spherical TPC detector filled with a high
pressure noble gas. An enhancement of the neutral current component is achieved via the coherent
effect of all neutrons in the target. Thus employing, e.g., Xe at $10$ Atm, with a feasible threshold energy
of about $100$ eV in the detection the recoiling nuclei,
 one expects between $600$ and $1900$ events, depending on the quenching factor.
We believe that networks of such dedicated detectors, made out of simple, robust and cheap technology,
 can be simply managed by an international scientific consortium and operated by students. This network
 comprises a system, which can be maintained
for several decades (or even centuries). This is   is a key point towards being able to observe
 few galactic supernova explosions.
 
acknowledgments: This work was supported in part by the
European Union under the contracts RTN No HPRN-CT-2000-00148 and
MRTN-CT-2004-503369.
%.......................................................................


\begin{thebibliography}{8.}
\addcontentsline{toc}{section}{References}

\bibitem{BEACFARVOG}
J.F. Beacom, W.M. Farr and P. Vogel, Phys. Rev D {\bf 66} (2002) 033001;hep-ph/0205220
\bibitem{SUPERNOVA}
J.R. Wilson and R.W.Mayle, Phys. Rept. {\bf 227} (1993) 97.\\
M. Herant, W. Benz, W.R. Hix, C.L. Fryer and S.A. Golgate, Astrophys. J. {\bf 435} (1994) 339.\\
M. Rampp and H.T. Janka, Astrophys. J. {\bf 539} (2000) L33.\\
A. Mezzacappa, M. Liebendorfer, O.E. Messer, W.R. Hix, F.K. Thielemann and S.W. Bruenn, Phys. Rev. Lett.
{\bf 86} (2001) 1935.\\
C.L. Fryer and A. Heger Astrophys. J. {\bf 541} (2000) 1033.\\
G.G. Raffelt, Nuc. Phys. Proc. Suppl. {\bf 110} (2002) 254;hep-ph/0201099;\\
R. Tomas, M. Kachellriess, G.G. Raffelt, 
A.Dighe, A-T Janka and L. Schreck, JCAP {\bf 0409} (2004) 015;\\
R. Tomas, D. Semikoz, G.G. Raffelt, M. Kachellriess and A.S. Dighe, Phys. Rev. D {\bf 68} (2002) 093013;\\
M.T. Keil, G.G. Raffelt, A-T Janka, Astrophys. J. {\bf 590} (2003) 971;\\
J.F. Beacom, R.N. Boyd and A. Mezzacappa, Phys. Rev. D {\bf 63} (2001) 073011.\\
M.K. Sharp, J.F. Beacom J.A. Formaggio, Phys. Rev. D {\bf 66} (2002) 013012; hep-ph/0205035.\\
A. Burrows, J. Hayes and B.A. Fryxell, Astrophys. J. {\bf 450} (1995) 830.
%\overline
%\bibitem{VERGADOS}
% See, e.g.\\
% J.D. Vergados, {\it  Phys. Rep.} {\bf 361} (2002) 1;\\
% J.D. Vergados, {\it  Phys. Rep.} {\bf 133} (1986) 1.
% \bibitem{NOSTOS1}
% Y. Giomataris and J.D. Vergados, Nucl. Instr. Meth. A {\bf 53} (2004) 330.
%\bibitem{VOGBEAC}
%P. Vogel and J.F. Beacom, {\it Phys. Rev. D} {\bf 60}
%(1999) 053003
\bibitem{SUPERKAMIOKANDE}
Y. Fukuda {\it et al}, The Super-Kamiokande  Collaboration,  {\it
Phys. Rev. Lett.} {\bf 86},  (2001) 5651; {\it ibid} {\bf 81}
(1998) 1562 $\&$ 1158; {\it ibid} {\bf 82}  (1999) 1810 ;{\it
ibid} {\bf 85} (2000) 3999.
\bibitem{SOLAROSC}
Q.R. Ahmad {\it et al}, The SNO Collaboration, {\it Phys. Rev.
Lett.} {\bf 89}  (2002) 011302; {\it ibid} {\bf 89}  (2002) 011301
;
{\it ibid} {\bf 87} (2001) 071301.\\
K. Lande {\it et al}, Homestake Collaboration, {\it Astrophys, J}
{\bf 496}, (1998) 505\\
W. Hampel {\it et al}, The Gallex Collaboration, {\it Phys. Lett.
B} {\bf 447}, (1999) 127;\\
J.N. Abdurashitov {\it al}, Sage Collaboration, {\it Phys. Rev. C}
{\bf 80} (1999) 056801;\\
G.L Fogli {\it et al}, {\it Phys. Rev. D} {\bf 66} (2002) 053010.
\bibitem{KAMLAND}
K. Eguchi {\it et al}, The KamLAND Collaboration, Phys. Rev. Lett.
90 (2003) 021802,  hep-exp/0212021.
 \bibitem{NOSTOS1}
 Y. Giomataris and J.D. Vergados, Nucl. Instr. Meth. A {\bf 53} (2004) 330.
 \bibitem{MICROPATTERN}
 P. Barbeau, J.I. Collar, J. Miyamoto and I. Shipsey, IEEE Trans. Nucl. Sci. {\bf 50} (2003) 1285.
 \bibitem{TWOPHASE}
 C. Hagmann and A. Bernstein, IEEE Trans. Nucl. Sci. {\bf 51} (2004) 2151.
 \bibitem{GIOMA96}
 I. Giomataris et al., Nucl. Instr. Meth. A {\bf 376} (1996) 29
 \bibitem{CG01}
J.I. Collar and Y. Giomataris, Nucl. Inst. Meth. {\bf 471} (2001) 254
\bibitem{GORO05}
 P. Gorodetzky et al., Nucl.Phys.Proc.Suppl. {\bf 138} (2005) 56
 \bibitem{WORKSHOP04}
  Spherical TPC Workshop, Paris 2004
  \bibitem{AALSETH}
   C.E. Aalseth et al., Nucl.Phys.Proc.Suppl. {\bf 110} (2002) 85.
   \bibitem{DARKMATTER}
 I. Giomataris et al., hep-ex/0502033.\\
 T. Patzak et al., Nucl. Instr. Meth. A {\bf 434} (1999) 358.\\
  B. Ahmed, G. J. Alner, H. Araujo, J. C. Barton, A. Bewick, M. J. Carson, D. Davidge, J. V. Dawson,
 T. Gamble, S. P. Hart, R. Hollingworth, A. S. Howard, W. G. Jones, M. K. Joshi, V. A. Kudryavtsev,
 T. B. Lawson, V. Lebedenko, M. J. Lehner, J. D. Lewin, P. K. Lightfoot, I. Liubarsky, R. Luscher, 
J. E. McMillan, B. Morgan, G. Nicklin, S. M. Paling, R. M. Preece, J. J. Quenby, J. W. Roberts, M.
 Robinson, N. J. T. Smith, P. F. Smith, N. J. C. Spooner, T. J. Sumner, D. R. Tovey,
Astropart.Phys. {\bf 19} (2003).
%\bibitem{BAHCALL02}
%J,N. Bahcall, M.C. Gonzalez-Garcia, C. Pe$\tilde{n}$a-Garay, {\it
%hep-ph/0212147}
%\bibitem{NUNOKAWA}
%H Nunokawa {\it et al}, {\it hep-ph/0212202}.
%\bibitem{ALIANI}
%P. Aliani {\it et al}, {\it hep-ph/0212212}.
%\bibitem{MSW}
%M. Maltoni, T. Schwetz and J.F. Valle, hep-ph/0212129;
%S. Pakvasa and J.F. Valle, hep-ph/0301061
%\bibitem{BARGER02}
%V. Barger and D. Marfatia, {\it hep-ph/0212126}.
%\bibitem{HOOFT}
%G. 't Hooft, Phys. Lett. B  {\bf 37} (1971) 195.
% \bibitem{REINES}
%F. Reines, H.S. Gurr and H.W. Sobel, Phys. Rev. Lett. {\bf 6}
%(1976) 315.
%\bibitem{ELNUNUC}
%S. Weinberg, Phys. Rev.D {\bf 5} (1972) 1412;\\ 
%L.A. Ahrens {\it et al}, Phys. Rev. D {\bf 35} (19b7) 785;\\
%C.H. Leewellyn Smith, Phys. Rep. {\bf 3} (1972) 261;\\
%S.M. Bilenky and J. Hosek, Phys. Rep. {\bf 90} (1982)73
\bibitem{VogEng}
P. Vogel and J. Engel, {\it Phys. Rev. D} {\bf 39} (1989) 3378.
\bibitem{PASCHOS}
E.A. Paschos and A. Kartavtsev, hep-ph/0309148.
\bibitem{SIMON03}
%E. Simon {\it et al}, SICANE:A detector Array for the Measurement
%of Nuclear Recoil Quenching Factors using a Monoenergetic Neutron
%Beam, astro-ph/0212491.\\
E. Simon {\it et al}, Nucl. Instr. Meth. A 507 (2003) 643; astro-ph/0212491.
\bibitem{LIDHART}
J. Lidhart {\it et al}, Mat. Phys. Medd. Dan. Vid. Selsk. 33 (10) (1963) 1
\bibitem{NOSTOS2}
The NOSTOS experiment and new trends in rare event detection,
I. Giomataris et al, 
hep-ex/0502033, submitted to the SIENA2004 International Conference (2005).
\bibitem{GRRC}
 Y. Giomataris, P. Rebourgeard, J.P. Robert, Georges Charpak,  Nucl. Instrum. Meth. A {\bf 376} (1996) 29
\end{thebibliography}
\end{document}